# THE DESIGN SCIENCE RESEARCH PROCESS: A MODEL FOR PRODUCING AND PRESENTING INFORMATION SYSTEMS RESEARCH

**Ken Peffers[1,2]**
University of Nevada, Las Vegas, College of Business Administration
4505 Maryland Parkway
Las Vegas NV 89154-6034 USA
Tel +1-702-807-1181, Fax +1-702-446-8370
k@peffers.com

**Tuure Tuunanen[2]**
The University of Auckland Business School, The Dept. of ISOM
Symonds Street 7, Private Bag 92019
Auckland, New Zealand
Tel +64-9-373-7599 ext. 84622, Fax: +64-9-373-7430
tuure@tuunanen.fi

**Charles E. Gengler**
City University of New York, Baruch College
17 Lexington Avenue, New York NY 10010 USA
Charles_Gengler@baruch.cuny.edu

**Matti Rossi**
Helsinki School Economics, Department of Business Technology
P. O. Box 1210, FIN-00101 Helsinki, Finland
matti.rossi@hse.fi

**Wendy Hui**
Hong Kong University of Science and Technology, Department of ISMT
Clear Water Bay, Kowloon, Hong Kong.
busywendy@gmail.com

**Ville Virtanen**
Helsinki School of Economics, LTT Research, Inc.
Pohjoinen Rautatiekatu 21 B, FIN-00100 Helsinki, Finland
ville.virtanen @hse.fi

**Johanna Bragge**
Helsinki School Economics, Department of Business Technology
P. O. Box 1210, FIN-00101 Helsinki, Finland
johanna.bragge@hse.fi

---

[1] Corresponding Author
[2] The two lead authors made substantially similar contributions to this paper. First authorship was determined by rotation among papers.

# THE DESIGN SCIENCE RESEARCH PROCESS: A MODEL FOR PRODUCING AND PRESENTING INFORMATION SYSTEMS RESEARCH

*Abstract*

The authors design and demonstrate a process for carrying out design science (DS) research in information systems and demonstrate use of the process to conduct research in two case studies. Several IS researchers have pioneered the acceptance of DS research in IS, but in the last 15 years little DS research has been done within the discipline. The lack of a generally accepted process for DS research in IS may have contributed to this problem. We sought to design a design science research process (DSRP) model that would meet three objectives: it would be consistent with prior literature, it would provide a nominal process model for doing DS research, and it would provide a mental model for presenting and appreciating DS research in IS. The process includes six steps: problem identification and motivation, objectives for a solution, design and development, evaluation, and communication. We demonstrated the process by using it in this study and by presenting two case studies, one in IS planning to develop application ideas for mobile financial services and another in requirements engineering to specify feature requirements for a self service advertising design and sales system intended for wide audience end users. The process effectively satisfies the three objectives and has the potential to help aid the acceptance of DS research in the IS discipline.

**Keywords**: Design science, design science research process, process model, case study, requirements engineering, requirements elicitation, and information systems development.

# INTRODUCTION

Information systems (IS) is an "applied" research discipline, we acknowledge, in the sense that we apply theory, frequently from other disciplines, such as economics, computer science, and the social sciences, to solve problems at the intersection of IT and organizations. Yet the dominant research paradigms we use to produce and publish research for our most respected research outlets largely continue to be traditional descriptive research borrowed from the social and natural sciences. More recently we've accepted interpretive research paradigms into our culture, but the resulting research output is still mostly explanatory and not often very applicable, it could be argued. While design, the act of creating an explicitly applicable solution to a problem, is an accepted research paradigm in other disciplines, such as engineering, this paradigm has been employed in just a small minority of research papers published in our best journals to produce artifacts of practical value to either the research or professional audience.

We wonder whether this may help to explain why the center of gravity for research in, for example, systems analysis and design, arguably IS research's *raison d'être,* seems to have moved to engineering, dominated by research streams like "requirements engineering" and "software engineering." Engineering disciplines accept design as a valid and valuable research methodology, but for the most part, major IS journals still seem to find it a questionable model for quality research. For example, a few years ago, when one of this paper's authors discussed the potential submission of an article about a better requirements analysis methodology with the editor-in-chief of one of the highest ranked IS research journals, he was told that the journal didn't entertain papers about new systems development methods, because they involved neither theory development nor theory testing.

In recent years several researchers have succeeded in bringing design research into the IS research community, successfully making the case for the validity and value of design science (DS) as an IS research paradigm (Hevner et al. 2004; March et al. 1995; Walls et al. 1992) and actually integrating design as major component of research (Nunamaker et al. 1991). In spite of these successful breaches, little DS research has been successfully published in the IS literature in the nearly 15 intervening years since those early papers (Walls et al. 2004).

What's missing may be a conceptual model for how researchers can carry out design science research in IS and a mental model or template for readers and reviewers to recognize and evaluate it. Every researcher trained in the culture of business research has mental models for empirical and theory building research (we know them instantly when we see them) and perhaps one for interpretive research as well. Even if all of these mental models aren't exactly the same; they provide contexts in which researchers can evaluate the work of others. For example, if a researcher reviewed an empirical paper that failed to describe how the data in was gathered, he or she would probably always regard that as an omission that required notice and correction. Since DS research isn't part of the dominant IS research culture, no such commonly understood mental model exists. Without one, how can DS research be evaluated or even distinguished from "mere" practice?

While a number of researchers, both in and outside of IS, have sought to provide some guidance to define design science research (Hevner et al. 2004) and to document the appropriate reference literature (Vaishnavi et al. 2005), so far no IS research has explicitly focused on the development of a conceptual process and mental model for carrying it out and presenting it. Such a process and mental model might help IS researchers to produce and present high quality design science research in IS that is accepted as valuable, rigorous, and publishable in the IS research literature.

In this paper our objective is to design a process for DS research in IS. Such a process should

> be consistent with design science processes in other disciplines,
>
> provide a nominal process for conducting the research, and
>
> provide a mental model for what DS research output looks like.

This research is expected to make several major contributions to IS researchers, the IS research publication community, and others.

For IS researchers, DS research offers an important paradigm for conducting applicable, yet rigorous, research, i.e., research that is closer to IS's applied *raison d'être*. A conceptual process could help researchers with a conceptual process for successfully carrying out DS research and a mental model for its presentation.

For the research publication community, including editors and reviewers, such a process and mental model could help them to recognize such research and to respect its objectives, processes, and outputs.

For the professional constituency for IS research, to the extent to which such a process encourages IS researchers to try to solve real problems with applicable solutions, the process may help to make some IS research more applicable and accessible.

Here we propose a conceptual process and mental model, the design science research process (DSRP) for the production and presentation of DS research in IS. We use the resulting process to develop the process and to present the results in this paper, i.e., the paper represents a demonstration of its own artifact. Consequently, the remaining sections of this paper are structured as follows. In subsequent sections,

> we review literature to understand the state of the problem (<u>problem identification</u>),
>
> we identify the objectives of a solution,
>
> we explicate the <u>design</u> of a DS research process,
>
> we demonstrate the use of the process in two case studies in which we develop methodology for use in IS planning and requirements engineering, where each of these case studies is also structured in accordance with the DS research process (<u>demonstration</u>), and lastly, we evaluate the process in terms of the objectives of our study (<u>evaluation</u>) and conclude.

## PROBLEM IDENTIFICATION: NEED FOR CONCEPTUAL PROCESS AND MENTAL MODEL

IS researchers have developed an interest in DS research over the last 15 years. Three papers from the early 1990s, (March et al. 1995; Nunamaker et al. 1991; Walls et al. 1992), introduced DS research to the IS community. March and Smith (1995) argued that design research could contribute by facilitating research to address the kinds of problems faced by IS practitioners. In their view design and natural science IS research are complementary to produce relevant and effective results for IS practice. Nunamaker et al. (1991) were interested in integrating system development into the research process. They proposed a multimethodological approach that would include 1) theory building, 2) systems development, 3) experimentation and 4) observations. All of these methodologies interact with each other and are essential for complete research products. Walls et al. (1992) took a more general approach to define information systems design theory, as a class of research that would stand as an equal with traditional social science based theory building and testing.

Once the methodological ice was broken for DS research, Walls et al. (2004) expected a rush to it in IS. DS research was the link between IS research and practice, and who wouldn't want IS research to have more impact on practice, to embed research in inventive systems, or to use design to develop better theory? For some reason, however, it doesn't seem to have happened. Walls et al (2004) lament that little DS research has been done in IS in the last 10 years. We collected the articles published in four top IS journals in the last 10 years in our own subject domain, requirements engineering (RE). RE seems a likely focus for DS research because of the domain's close links with practice and because the outcome of research in the domain is often a new methodology, a potential artifact in design research. Of the 15 articles, just two could arguably be considered DS research (see Appendix I for details). Given that many software engineering or computer science papers take such a design science approach (Morrison et al. 1995), we wonder why it shouldn't be happening in IS.

One reason why DS research hasn't been widely adopted in IS might be a lack of a conceptual process for carrying it out and a mental model for presenting the research and evaluating the outputs. Such a process and model could help researchers produce rigorous research acceptable for the IS community and help reviewers and the researcher audience recognize it as such.

Other research traditions have addressed these issues. Engineering, for example, appears to use a wide variety of approaches to conduct design research (Evbuonwan et al. 1996), i.e., research about engineering design. The emphasis in engineering has been quite applied and the focus could be described as process targeting the production of artifacts (van Aken 2005). Evbuonwan et al (1996) mention fourteen different process models. Many of these, like Cooper's StageGate (Cooper 2000; Cooper 1990), are clearly intended more as design or development methodologies than research methodologies. Since DS has strong roots in design it is not surprising that many of papers concentrate more on doing research about design (Eekels et al. 1991; Hybs et al. 1992; Macmillan et al. 2001).

Some engineering researchers, e.g., (Eekels et al. 1991), have seen a need a for common DS research methodology,. For example, Archer's (1984) methodology focuses on one kind of DS research, that which resulted in building system instantiations as the research outcome. Archer's own research the research outcomes included better designs for hospital beds and gadgets that prevented fire doors from being propped open. He defined DS research in six steps: programming (to establish project objectives), data collection and analysis phases, synthesis of the objectives and analysis results, development (to produce better design proposals), prototyping, and documentation (to communicate the results).

In spite of efforts to produce DS research guidance (Fulcher et al. 1996; Reich 1994), there is still no consensus on what it should include and design science research still lacks a shared methodology (Fulcher et al. 1996) that would provide a general process for IS DS research.

IS researchers have addressed the issue of what goals to pursue in producing DS research, e.g. (Fulcher et al. 1996; Hevner et al. 2004), and they have worked to provide theoretical frameworks to justify design research studies, e.g. (Adams et al. 2004; Nunamaker et al. 1991; Walls et al. 1992). However, there haven't been efforts to explicitly focus on the development of a consensus process and mental model for such research that engineering researchers have called for (Fulcher et al. 1996; Reich 1994). This might help to explain Walls et al's (2004) findings about how their 1992 article had impacted the IS community; twenty-six papers cite their paper, but their overall feeling was that the message had not gotten through.

## OBJECTIVES OF A SOLUTION: PROCESS AND MENTAL MODEL CONSISTENT WITH PRIOR RESEARCH

In the introduction of the paper we defined our objective for the paper as the development of a conceptual process for design science research in IS and mental model for its presentation. We suggested that the process model should be consistent with design science processes in IS and other disciplines, should provide a nominal process for conducting the research, and it should provide a mental model for what DS research output looks like.

The prior literature provides us with many explicit process models in engineering for design research, i.e., process models for engineering design. It also provides us with a few implicit models for DS research in IS. If our design science research process (DSRP) is consistent with prior literature, this will provide us with a *prime facie* case that it is serviceable and flexible enough to support DS research in IS. Thus, consistency with prior research is our first objective.

There are a few papers that are particularly useful to address the problem of a DSRP model. We found Nunamaker, Chen, and Purdin's (1991) system development research methodology to be interesting for this purpose. Walls, Widmeyer, and El Sawy's (1992; 2004) information system design theory is similarly helpful. In engineering, Archer (1984) and Eekels and Roozenburg (1991) have presented process models for design that have synergies with the IS originated models. Additionally, March and Smith's (1995) and Hevner et al.'s (2004) guidelines for design science research influence methodological choices within the DS research process. In the computer science domain, Takeda et al. (1990) proposed a "design cycle" for intelligent design systems. Two informal publications in the current IS literature also contribute something to the conversation. In a workshop presentation, Rossi and Sein (2003) suggested basic DS research steps. Vaishnavi and Kuechler's (2005) ISWORLD website, "Design Research in Information Systems," provides rich information about the historical context for design research.

The second objective is that such a process should provide a nominal process for conducting DS research in IS. Other research paradigms, such as theory testing or interpretive research, that are well accepted in IS research culture, have well defined nominal processes that are reflected in research articles on research practice, PhD training, and the examples of many research efforts, published in prominent journals and prize winning dissertations, that are acknowledged to be well done. A nominal process could not be a literal, sequential procedure because circumstances, opportunities, and resource constraints effect how researchers proceed.

The third objective of a DSRP model is to provide a mental model for the characteristics of research outputs. Outcomes from DS research are clearly expected to differ from those of theory testing or interpretation and a DSRP model should provide us with some guidance, as reviewers, editors, and consumers, about what to expect from DS research outputs.

In the next section we use prior research about design and design science research to design a DS research process.

## DESIGN: DEVELOPMENT OF THE PROCESS

To design our DSRP, we looked to influential prior research and current thought to determine what common process elements could be found in the literature. In Table 1 we present process elements from seven representative papers and presentations and our synthesis: the components of an IS design science research process. Note that when we look at Table 1 we should not lose sight of the fact that these researchers had distinctly different purposes in mind, so we're using them to draw out a consensus of common elements, rather than trying to compare "apples to oranges," to quote a helpful reviewer. We can see that the authors agree substantially on common elements. All seven include some component in the initial stages of research to define a research problem. Nunamaker et al. (1991) and Wall et al. (1992) emphasize theoretical bases, while engineering researchers (Archer 1984; Eekels et al. 1991) take a more practical approach to suggest problem analysis. Takeda et al (1990) and Rossi and Sein (2003) also include problem identification as a basic process step, as do Hevner et al (2004) in their general guidelines for DS research. Consequently, we include *problem identification and motivation* in our synthesis. Some of the researchers explicitly incorporate efforts to transform the problem into system objectives, also called meta-requirements (Walls et al. 1992) or requirements (Eekels et al. 1991), while for the others this effort is implicit, e.g., part of programming and data collection (Archer 1984). For Hevner et al (2004), in turn, this is implicit in their relevance guideline. Our synthesis for this is *objectives of a solution.*

All of the researchers focus on the core of design science across disciplines: *design and development*. In some of the research, e.g., (Eekels et al. 1991; Nunamaker et al. 1991) the design and development activities are further subdivided into more discrete activities whereas other researchers discuss more about the nature of the iterative search process (Hevner et al. 2004). Next, the solutions vary from single act of *demonstration* (Walls et al. 1992) to prove the idea works to more formal *evaluation* (Eekels et al. 1991; Hevner et al. 2004; Nunamaker et al. 1991; Rossi et al. 2003; Vaishnavi et al. 2005) of the developed artifact. Eekels et al. (1991) and Nunamaker et al. (1991) include both of these phases. Finally, Archer (1984) and Hevner et al (2004) proposed the need of *communication* to diffuse the resulting knowledge.

The components in our synthesis of this prior research and their sequential order form the basis for our design science research process.

The result of our synthesis is a process model consisting of six activities in a nominal sequence, here described and presented graphically in Figure 1.

*1. Problem identification and motivation.* Define the specific research problem and justify the value of a solution. Since the problem definition will be used to develop an effective artifactual solution, it may be useful to atomize the problem conceptually so that the solution can capture the problem's complexity. Justifying the value of a solution accomplishes two things: it motivates the researcher and the audience of the research to pursue the solution and to accept the results and it helps to understand the reasoning associated with the researcher's understanding of the problem. Resources required for this activity include knowledge of the state of the problem and the importance of its solution.

*2. Objectives of a solution.* Infer the objectives of a solution from the problem definition. The objectives can be quantitative, e.g., terms in which a desirable solution would be better than current ones, or qualitative, e.g., where a new artifact is expected to support solutions to problems not hitherto addressed. The objectives should be inferred rationally from the problem specification. Resources required for this include knowledge of the state of problems and current solutions and their efficacy, if any.

*3. Design and development.* Create the artifactual solution. Such artifacts are potentially, with each defined broadly, constructs, models, methods, or instantiations (Hevner et al. 2004). This activity includes determining the artifact's desired functionality and its architecture and then creating the actual artifact. Resources required moving from objectives to design and development include knowledge of theory that can be brought to bear as a solution.

*4. Demonstration.* Demonstrate the efficacy of the artifact to solve the problem. This could involve its use in experimentation, simulation, a case study, proof, or other appropriate activity. Resources required for the demonstration include effective knowledge of how to use the artifact to solve the problem.

**Table 1 Design and design science process elements from IS other disciplines and synthesis objectives for a design science research process in IS.**

| Objectives for a design science research process model | Archer (1984) | (Takeda et al. 1990) | Eekels and Roozenburg (1991) | Nunamaker et al (1991) | Walls et al (1992) | (Rossi et al. 2003) | (Hevner et al. 2004) |
|---|---|---|---|---|---|---|---|
| **1. Problem identification and motivation** | Programming Data collection | Problem enumeration | Analysis | Construct a conceptual framework | Meta-requirements Kernel theories | Identify a need | Important and relevant problems |
| **2. Objectives of a solution** | | | Requirements | | | | Implicit in "relevance" |
| **3. Design and development** | Analysis Synthesis Development | Suggestion Development | Synthesis, Tentative design proposals | Develop a system architecture Analyze and design the system. Build the system | Design method Meta design | Build | Iterative search process Artifact |
| **4. Demonstration** | | | Simulation, Conditional prediction | Experiment, observe, and evaluate the system | | | |
| **5. Evaluation** | | Confirmatory evaluation | Evaluation, Decision, Definite design | | Testable design process/product hypotheses | Evaluate | Evaluate |
| **6. Communication** | Communication | | | | | | Communication |

*5. Evaluation.* Observe and measure how well the artifact supports a solution to the problem. This activity involves comparing the objectives of a solution to actual observed results from use of the artifact in the demonstration. It requires knowledge of relevant metrics and analysis techniques. Depending on the nature of the problem venue and the artifact, evaluation could include such items as a comparison of the artifact's functionality with the solution objectives from activity 2 above, objective quantitative performance measures, such as budgets or items produced satisfaction surveys, client feedback, or simulations. At the end of this activity the researchers can decide whether to iterate back to step 3 to try to improve the effectiveness of the artifact or to continue on to communication and leave further improvement to subsequent projects. The nature of the research venue may dictate whether such iteration is feasible or not.

*6. Communication.* Communicate the problem and its importance, the artifact, its utility and novelty, the rigor of its design, and its effectiveness to researchers and other relevant audiences, such as practicing professionals, when appropriate. In scholarly research publications researchers might use the structure of this process to structure the paper, just as the nominal structure of an empirical research process (problem definition, literature review, hypothesis development, data collection, analysis, results, discussion, and conclusion) is a common structure for empirical research papers. Communication requires knowledge of the disciplinary culture.

This process is structured in a nominally sequential order; however there is no expectation that researcher(s) would always actually proceed in sequential order from activity 1 through activity 6. Instead they may actually start at almost any step and move outward. A problem-centered approach is the basis of the nominal sequence, starting with activity 1. Researchers might proceed in this sequence if the idea for the research resulted from observation of the problem or from suggested future research in a paper from a prior project. An objective centered solution, starting with activity 2, could be the by-product of consulting experiences, where, for example, the results of system development activities that fell short of hopes and clients wished that we could do a better job of scheduling offshore programming. A design and development centered approach would start with activity 3. It would result from the existence of an artifact that has not yet been formally thought through as a solution for the explicit problem domain in which it will be used. Such an artifact might have come from another research domain, it might have already been used to solve a different problem, or it might have appeared as an analogical idea. Finally, observing a practical solution that worked, starting with activity 4, might result in a design science solution if researchers work backwards to apply rigor to the process retroactively.

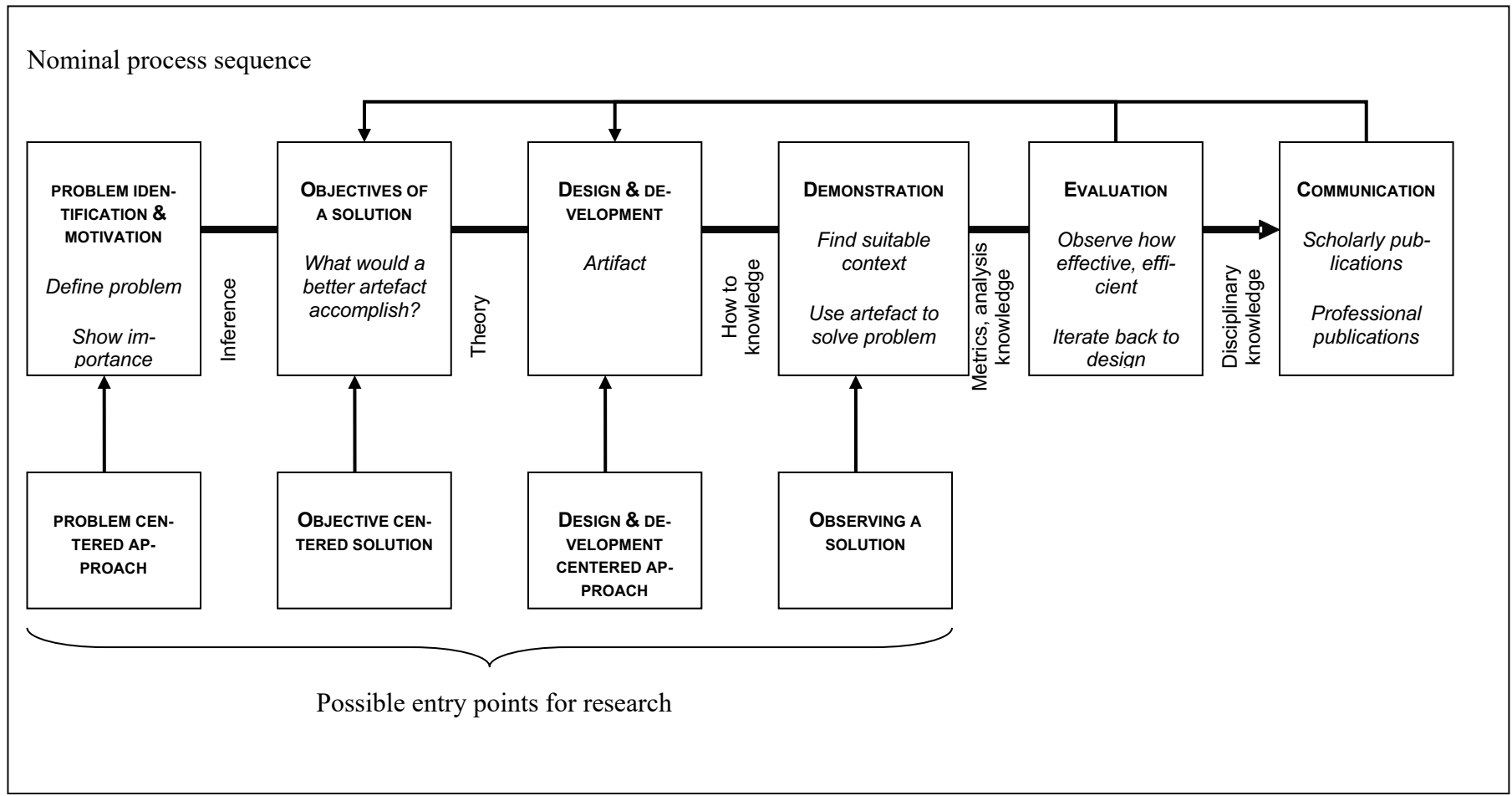

**Figure 1 Design science research process (DSRP) model**

# DEMONSTRATION IN TWO CASE STUDIES

Here we demonstrate the use of the DSRP in two ways. First, the development of the DSRP model is a case of the use of the process itself. Secondly, we tell the story, below, of how we use the DSRP in two cases, one in IS planning and the other in requirements engineering.

The development of the DSRP model included six steps, which are echoed in the sections of this paper.

**Problem identification and motivation**. This included our recognition that IS research lacked a process model for DS research in IS and the effect of this on the adoption of DS research as a methodology in IS.

**Objectives of a solution**. The objectives included consistency with DS processes in prior literature, the development of a nominal process for conducting research, and the development of a mental model for DS research output.

**Design and development**. We used prior literature as the basis for the development of a six-step process model.

**Demonstration**. This involved demonstrating the resulting process model with case studies.

**Evaluation**. This will involve comparing the characteristics of the DS research process model with the objectives described above.

**Communication**. Communication is being done through this paper.

Here we further demonstrate the use of the DSRP through two case studies in which we have used it to develop artifacts. The first one depicts a critical success chains (CSC) method (Peffers et al. 2003a; Peffers et al. 2005a) to develop a portfolio of valuable applications for a new technology and the second one describes the development of the wide audience requirements engineering (WARE) method (Tuunanen et al. 2004) in the *Helsingin Sanomat* case. In each case the developed artifact was the method, not the applications even though these were well-received "by-products" by the industry clients. We provide case descriptions of each and illustrate how the studies have used the DSRP.

## *Digia: developing a method to generate ideas for new applications that customers value*

SysOpen Digia Plc, is a Helsinki based research and development firm specializing in innovative software applications for wireless communication that focuses on the creation of personal communication technologies and applications for the next generation wireless information devices, such as Smart phones and Communicators. It employs approximately 800 professionals and realized revenue of approximately € 64 million in 2004. This is the story of our efforts at Digia, also illustrated in Figure 2 and reported in detail in (Peffers et al. 2003a; Peffers et al. 2005b), to develop a better IS planning method.

**Objective centered solution**. In fall 2000, Digia chairman, Pekka Sivonen, approached us with a request to help define a portfolio of potential applications for Digia to develop to meet the need for financial services delivered by the next generation wireless devices (Peffers et al. 2003a). We accepted his invitation because it fit our current research objective: to develop a method to support the generation of ideas for IS projects that would provide the greatest impact on achieving a firm's strategic goals. Since applications for providing financial services using mobile devices hadn't much been done before, this looked like a good opportunity to use a new method to develop a portfolio of applications using a new planning method.

**Problem identification and motivation.** In most firms there is no shortage of ideas for new IS projects, but most tend to be suboptimal. How can managers make use of the ideas of many people within and around the organization, while keeping the focus on what is important and valuable for the firm? Bottom up planning generates many ideas, but most are self serving and narrowly focused. Top down planning ignores the ideas of all except those in the executive suites.

**Objectives of a solution**. Our objective was to develop an IS planning method that allowed us to make use of the ideas of many, including experts outside the firm and potential users, but to keep the focus on ideas with high strategic value to the firm.

**Design & development**. We used personal construct theory (PCT) (Kelly 1955) and critical success factors (Rockart 1979) as theoretical bases for the method development. For data collection, we used "laddering," a PCT based technique for structured interviewing that collects rich data on subject reasoning, as well as preferences. For analysis we adapted hierarchical value maps, which had been used in marketing to display aggregated laddering data graphically. We incorporated an ideation workshop, where business and technical expertise was brought to bear on the task of developing feasible ideas for new business applications from the graphical presented preferences and reasoning of the subjects. The result of the design effort was the critical success chain (CSC) method for using the ideas of many people in and around the organization to develop portfolios of feasible application ideas that are highly valuable to the organization.

**Demonstration**. We used the opportunity at Digia to demonstrate CSC's feasibility and efficacy (Peffers et al. 2003a; Peffers et al. 2005a). We started by recruiting and interviewing 32 participants, approximately evenly divided between experts and potential end users. We conducted individual structured interviews, using stimuli collected from the subjects ahead of time. The interview method was intended to encourage participants to focus on the value of ideas.

The laddering interviews provided us with rich data about applications the participants wanted and why. Using qualitative clustering, we used the data to create five graphical maps, containing 114 preference and reasoning constructs. The next step was to conduct an ideation workshop with six business and engineering experts and managers in the firm to convert the participant preferences to feasible business project ideas at a "back-of-the-envelop" level. In the workshop, conducted in isolation in a single five-hour stretch, the participants developed three business ideas, with application descriptions, business models, and interaction tables. These were further developed by analysts in post workshop work to be integrated into the firm's strategic planning effort.

**Evaluation**. Digia representatives were very enthusiastic about the results of the workshop. After the workshop, Digia's chairman remarked that the workshop "positively . . . exceeded [his] expectations [about] the results" (Peffers et al. 2003a). This feedback with the successful implementation of the method in practice enabled us to present initial "proof-of-concept" level validation of our method (Peffers et al. 2003a; Peffers et al. 2005a). The firm intended to use the resulting applications to plan its continued product development efforts.

**Communication**. The case study was successfully reported in reputable scholarly journals, including *Journal of Management Information Systems* (Peffers et al. 2003b) and *Information & Management* (Peffers et al. 2005c). Furthermore, the findings have been presented in numerous practitioner oriented outlets, such as book chapters, e.g., (Peffers et al. 2005c), technical reports, e.g., (Tuunanen 2001), and trade magazine articles, e.g., (Tuunanen 2002).

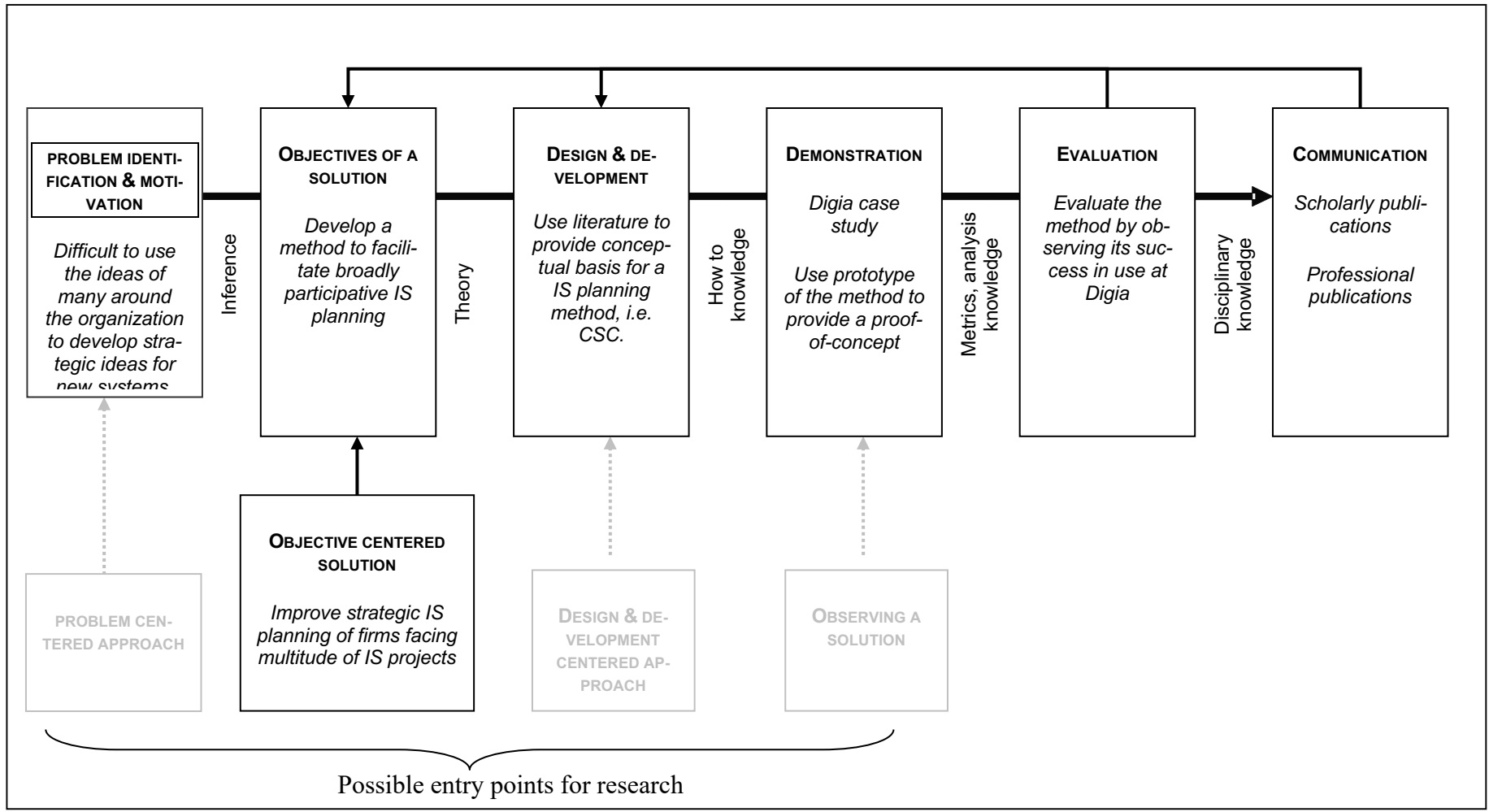

**Figure 2 DS research process for the Digia study.**

*Helsingin Sanomat: developing a method for requirements engineering systems for wide audience end users*

*Helsingin Sanomat* is the one of the biggest daily newspapers in the Nordic countries with more than one million daily readers. It is also Finland's leading advertising medium, with more than 400 thousand ads printed yearly. We were engaged by their business development team to develop the functional requirements for the Medianetti e-Ad Traffic and Ad Information System (META-IS), version 2.0 and beyond. META-IS is a system for self-service sales and design of newspaper display advertising for *Helsingin Sanomat* and several sister publications. The case is illustrated in terms of the DS research process Figure 3 and it has been report more in detail in (Tuunanen et al. 2004).

**Problem centered approach**. We wanted to develop a better method to understand the feature requirements for systems that would be used by wide audience end users, i.e., users who were outside the organization, widely dispersed, with little connection to the organization and little motivation to participate in requirements analysis, and perhaps with little understanding of the underlying technology and business models. Requirements engineering for META-IS version 2 provided us with an opportunity to study this problem because many of the intended users for the system, infrequent advertisers and small businesses, were naïve advertisers with little connection to the firm.

**Problem identification & motivation**. We recognized the underlying problem of RE for wide audience end users while we were working at Digia on the IS planning method (Tuunanen 2003). We came to understand that more and more, the IS community was facing the problem of developing systems positioned for end-users outside of the traditional environment, the firm or organization. Peffers and Tuunanen (2005a) had identified the important characteristics of wide audience end-users and the needs they create for a requirements elicitation method (Tuunanen et al. 2004).

**Objectives of a solution.** We wanted to develop a method that would facilitate understand user preferences, values, and reasoning for features for systems to be used by such users. Such a method would allow users with no connection to the firm and little understanding of the underlying business or technology to participate in feature level planning of new applications.

**Design & development.** The development of the new method adapted and extended concepts that we had developed for IS planning in the CSC method. We used literature to provide conceptual bases for a wide audience requirements engineering (WARE) method (Tuunanen et al. 2004). We addressed issues identified earlier, such as how to reach widely dispersed end-users or how to avoid problems arriving from lack of context awareness of end-users of the developed IS artifact, by using our earlier experiences with the Digia case (Peffers et al. 2003a; Peffers et al. 2005a) and extending the theory base through the use of relevant literature. This effort resulted to new conceptual requirements engineering method.

**Demonstration**. We tested WARE by using it to develop features and a product upgrade roadmap for META-IS version 2. META-IS allows customers to purchase and design display advertising. It is targeted to serve five customer segments, regular and infrequent small-scale advertisers, medium scale advertisers, large-scale advertisers, the media and ad agencies, and internal users in the firm. Hence, the end-users are very diverse in their backgrounds and needs. Furthermore, they are dispersed throughout Finland and the end-user potential is calculated in tens of thousands or even hundreds of thousands, in short nearly every firm or organization with adverting needs in Finland.

We selected 30 participants from different end-users segments to participate in the study, focusing on potential lead users (von Hippel 1986). We used the laddering interview technique, with categorical stimuli provided by the firm, to collect data that included 2566 distinct feature preferences and reasons, which were later clustered to five themes. We used these themes to create graphical maps and developed a presentation tool that would allow developers and managers to drill down from the maps to see individual participant reasoning and even to listen to representative subject interview segments. A business and technical workshop and additional analysis resulted in a business report and a three-year upgrade roadmap for the application.

**Evaluation**. The evaluation of WARE's efficacy at *Helsingin Sanomat*, was threefold. First, we set up a workshop with the client's development team to help them learn to use the presentation tool and to assess the results. Secondly, the results of the workshop provided us with feedback towards planning the next step: use of a survey of potential users to evaluate the theme maps, i.e., to determine which features were most valuable for the users. Thirdly, we had feedback from the client on the value of the resulting business report and application development roadmap. This report was used extensively within the organization and was praised by the development team's manager. The client's actual development roadmap was nearly entirely based on features developed in the study and it has subsequently been largely implemented (Tuunanen et al. 2004).

**Communication**. The case study is currently under its way towards scholarly publications with first one targeted toward an IS research journal (Tuunanen et al. 2004). Similarly, the findings are starting find their way towards practitioner outlets through book chapters, e.g., (Tuunanen 2005), and print and TV media interviews.

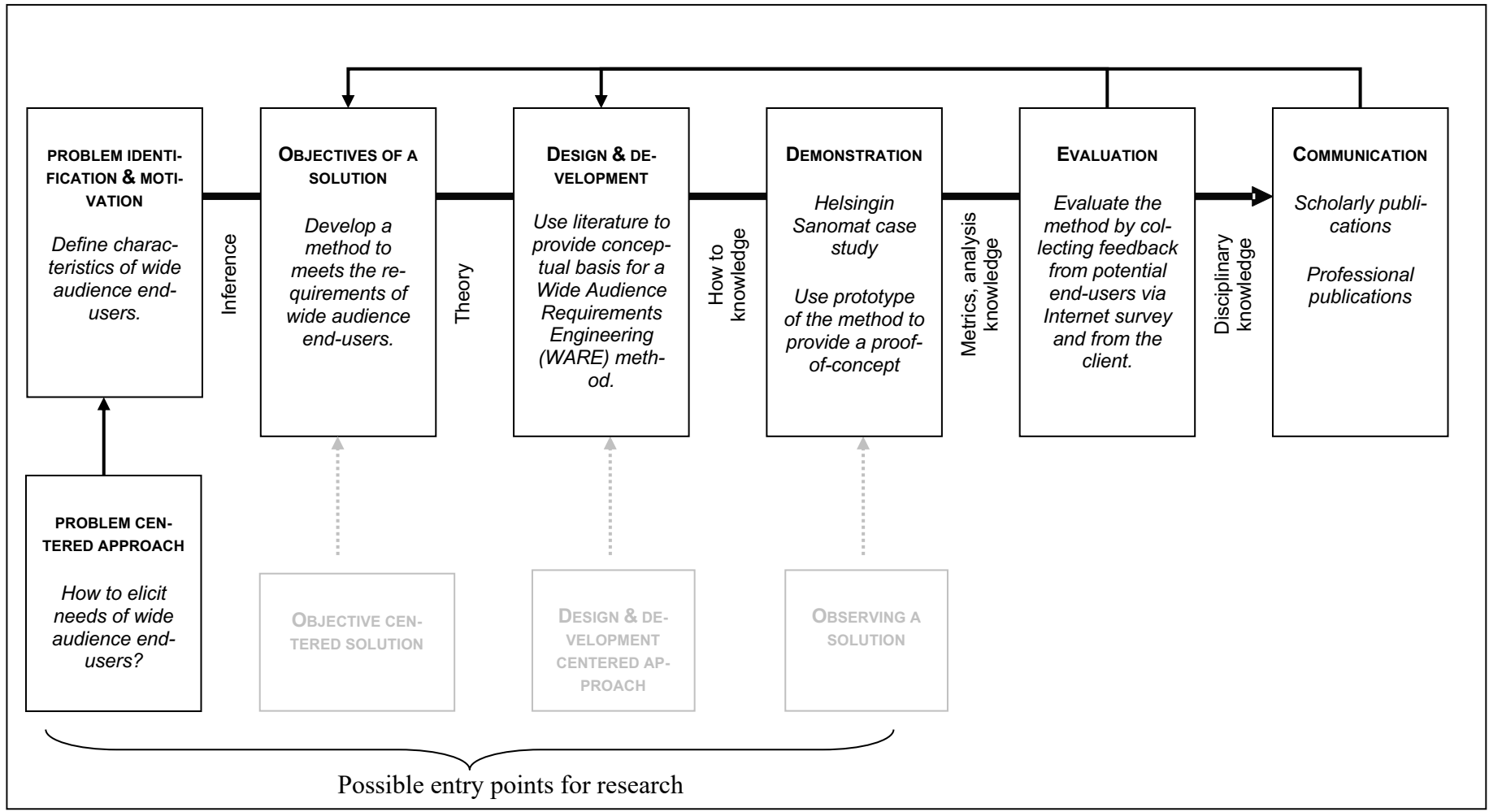

**Figure 3 DS research process for the *Helsingin Sanomat* study.**

## Evaluation of the DSRP model

We evaluate the DSRP model in terms of the three objectives for DSRP described above. First, is it consistent with prior DS research theory and practice, as it has been represented in the IS literature, and with design and design science research, as it has been conveyed in representative literature in reference disciplines? Second, does it provide a nominal process for conducting DS research in IS? Third, does it provide a mental model for the characteristics of research outputs?

First, the DSRP model is consistent with concepts in prior literature about design science in IS. We can verify this by referring back to our analysis in Table 1 of process elements in the prior literature. Nunamaker, Chen, and Purdin's (1991) system development research methodology is far more general in its objectives than the DS process even though their methodology is focused on system building activities, rather than research processes, its five-step methodology can be mapped roughly to the DS process. Likewise Walls, Widmeyer, and El Sawy's (1992; 2004) "components of an information system design theory," Takeda et al's (1990) "design cycle" solution for intelligent computer aided design systems, Rossi and Sein's (2003) steps, Archer's (1984) process for industrial design, Eekels and Roozenburg's (1991) process for engineering design, and Hevner et al's (2004) guidelines for the required elements of design research, are all consistent with the DSRP.

Secondly, the DSRP model provides a nominal process for conducting DS research. We've used this process on four projects so far, including the development of the DSRP itself, i.e., this paper, the two case studies presented in the demonstration section of this paper, and a fourth on-going research project focusing on the problem of developing culturally segmented feature sets for applications to be used by global wide audience end-users. In each case we've conducted the project using a problem or objective centered approach, the process worked well, and it was effective for its intended purpose.

*Thirdly*, The DSRP also provides a mental model for the presentation of research outputs from DS research. In each of the four projects, including this one, we've used the DSRP model as the template for creating research publication outputs, such as this paper. In two cases, so far (Peffers et al. 2003c; Peffers et al. 2005a), these papers have been accepted for publication in well-ranked IS research journals, suggesting that the DSRP model serves as an effective presentation template. We expect that this will also prove true with the subsequent papers and future papers that might use it as a presentation template.

## CONCLUSION

In this paper we sought to develop a conceptual process for DS research in IS. We wanted this process model to be in harmony with design science processes in IS and in other related disciplines. Furthermore, we wanted the process model to guide researchers who would work on DS research and provide a mental model for what parts of DS research output might look like.

Our study contributes by providing a conceptual DS research process model that addresses the limitations of the literature. Until now there hasn't been an explicit process model intended to guide researchers who want a roadmap for how to do design science research in information systems. The DSRP model fills this role with a process that is consistent with prior literature and, for that reason, likely to be found sufficiently complete and robust to serve as an effective template for future research.

Another contribution of this research is to provide a mental model for what design science research in IS should look like. In the six steps of the process and in the look of this paper and its successors, researchers, reviewers, editors, and the design science research audience should find the kind of template that can help design science research become a part of the IS research culture. Just as researchers have a mental model for theory testing research and what components that they expect publications from it, i.e., research problem, hypotheses, data collection, analysis, results, discussion, etc., they'll have an analogous mental model for DS research outputs.

We hope that many DS researchers in IS will try to use this process model, thereby testing its usability. The case studies we have provided with this paper demonstrate its use, but only within the scope of two research problems. Further use will tell us whether there are problem domains where it requires extension or where it just doesn't work well. Another interesting problem is that of the research entry point. We made a point of suggesting that there might be multiple possible entry points for design science research. Of course, this issue isn't unique to design science research. We don't recall reading a theory testing paper where the authors say that they decided on the research questions after they collected the data or even after they did the analysis, but we guess that surely this happens with no ill effects. We think that a research process model should account for it. Wouldn't it be refreshing to read an academic paper where the researchers admitted that they had designed an artifact and then found a problem that it solved? How would that affect the design process and the design itself?

## APPENDIX 1: KEYWORD SEARCH PROCEDURE

*Objective*: Determine how many DS research papers have been published in the specific domain of our own study, requirements engineering.

*Keywords used*: By consensus among the authors, we used the keywords in the search; "requirements elicitation" and "requirements engineering," as keywords commonly used within software engineering and computer science, and "requirements analysis" and "requirements determination," terms that were thought to be more prevalently used among IS researchers.

*Search details*: All four searches done 27$^{th}$ September 2005 and they produced in total of 840 references. Eliminating duplicate publications reduced the number to 768. We refined the search by using a published journal ranking list (Peffers et al. 2003d), and to the top-ranked four journals, *Journal of Management Information Systems*, *MIS Quarterly*, *Information Systems Research*, and *Information & Management*. This left us with 15 articles. For details please see Table 2 in below.

**Table 2 The conducted structured literature review**

| Step | Outcome |
|---|---|
| Define key words for ISI Web of Science topic search | Requirements Analysis, Requirements Determination, Requirements Elicitation, and Requirements Engineering |
| Search "Requirements Analysis" | Found 338 references |
| Search "Requirements Determination" | Found 46 references |
| Search "Requirements Elici- | Found 99 references |
| tation" | |
| Search "Requirements Engineering" | Found 357 references |
| Erasing duplicates | 768 references |
| Target search to main stream and high quality IS journals | *Journal of Management Information Systems (JMIS), MIS Quarterly (MISQ), Information Systems Research (ISR)* and *Information & Management (I&M).* |
| Limit search to I&M, ISR, JMIS and MISQ | 15 references |
| Methodology analysis | Please see Table 3 |

*Final results of the search*: We analyzed each of the 15 articles qualitatively to determine the dominant research methodology it used. These results are presented in Table 3 in below, namely we found two articles that can be said to use design science as a methodology.

**Table 3 Requirements engineering studies published in past ten years in main stream IS journals (I&M, ISR, JMIS & MISQ)**

| Article | Theory testing | Interpretive / Explorative | Conceptual / review | Design science |
|---|---|---|---|---|
| 1. Tiwana et al. (2005) | √ | | | |
| 2. Peffers et al. (2005a) | | | | √ |
| 3. Pitts et al. (2004) | √ | | | |
| 4. Hickey et al. (2004) | | | √ | |
| 5. Zoryk-Scahlla et al. (2004) | | √ | | |
| 6. McKinney et al. (2004) | √ | | | |
| 7. Duggan et al. (2004) | √ | | | |
| 8. Diaz et al. (2004) | | | | √ |
| 9. Davidson (2002) | | √ | | |
| 10. Browne et al. (2002) | | | √ | |
| 11. Browne et al. (2001) | √ | | | |
| 12. Eva (2001) | | | √ | |
| 13. Scott et al. (2000) | | √ | | |
| 14. Marakas et al. (1998) | √ | | | |
| 15. Purvis et al. (1997) | √ | | | |